\documentstyle[12pt,aasms4]{article} 
\newcommand{\hmp}{h^{-1}Mpc}

\newcommand{\be}{\begin{equation}} 
\newcommand{\ee}{\end{equation}} 
\newcommand{\bea}{\begin{eqnarray}} 
\newcommand{\eea}{\end{eqnarray}} 
 
\newcommand{\bef}{\begin{figure}} 
\newcommand{\eef}{\end{figure}}

\def\spose#1{\hbox to 0pt{#1\hss}}  
\def\ltapprox{\mathrel{\spose{\lower 3pt\hbox{$\mathchar"218$}}  
 \raise 2.0pt\hbox{$\mathchar"13C$}}}  
\def\gtapprox{\mathrel{\spose{\lower 3pt\hbox{$\mathchar"218$}}  
 \raise 2.0pt\hbox{$\mathchar"13E$}}}  
\def\inapprox{\mathrel{\spose{\lower 3pt\hbox{$\mathchar"218$}}  
 \raise 2.0pt\hbox{$\mathchar"232$}}}  
  
\slugcomment{Subimitted to Astrophys. J. Letters} 
 
\lefthead{Sylos Labini \& Gabrielli} 
\righthead{Persistent fluctuations  
and scaling properties in galaxy number counts}

\begin{document}  
\title {Persistent fluctuations  
and scaling properties in galaxy number counts} 
 
\author{Francesco Sylos Labini \altaffilmark{1,3} 
and  Andrea Gabrielli\altaffilmark{2,3} }

\altaffiltext{1}{D\'epartement de Physique Th\'eorique,  
Universit\'e de Gen\`eve, 
24 quai Ernest Ansermet, CH-1211 Gen\`eve 4, Switzerland}

\altaffiltext{2}{\'Ecole Polytechnique, 91128 - Palaiseau Cedex, 
 France} 
 
\altaffiltext{3}{INFM Sezione Roma1,         
		      Dip. di Fisica, Universit\'a "La Sapienza",  
		      P.le A. Moro, 2,   
        	      I-00185 Roma, Italy. }

\setcounter{footnote}{0} 
 
\begin{abstract}  
 
Counts of galaxies as a function of apparent 
magnitude are among the most time-honored 
observations in cosmology. 
In this Letter,  
we focus on some statistical properties of these 
counts which are 
fundamental in order to characterize the large scale 
correlations in the galaxy spatial distribution. 
There are, in fact, no longer doubts since  
two  decades  
of the existence of very large 
scale structures. The  still remaining problem
concerns the correct characterization of their 
statistical properties.  
We propose to study  two properties of
galaxy counts data, 
in order to discriminate  
between a small scale ($\sim 5 \div 20 \hmp$) homogeneous 
distribution, and  a fractal structure on large scales 
($\sim 20 \div 300 \hmp$). 
Firstly, the average slope of the counts  which 
can be associated to an eventual fractal dimension in real  
space by simple arguments. 
Note that in the magnitude range
$11^m \ltapprox B_J \ltapprox 18^m$, 
the results are nearly independent  
from cosmological parameters, K-corrections and evolution 
effects, as the corresponding average redshift is  
$\langle z \rangle \ll 0.1$. 
Secondly, we propose to study  fluctuations of counts 
around the average behavior as a function of apparent magnitude 
in the whole magnitude range $B_j \gtapprox 11^m$. 
These fluctuations can  
discriminate  between a genuine fractal distribution 
and a homogeneous one.
In fact, they are related to the very 
statistical properties 
of the spatial distribution, independently on cosmological corrections. 
More specifically, in a fractal distribution 
one expects to find persistent scale-invariant fluctuations
around the average behavior, 
which do not decay with apparent magnitude. 
On the other hand, in an homogeneous distribution,
on large enough scales, 
the relative 
variance of the counts should decrease exponentially with  
apparent magnitude.  
Smooth cosmological corrections cannot change such a behavior. 
Such a test  can be applicable also at very faint magnitudes. 
We point out that the application of these tests 
to the new generation of  
photometric (POSS-II) 
and spectroscopic  
surveys (SDSS, 2dF) will be crucial 
in order to characterize statistically the galaxy spatial distribution. 
\end{abstract} 
 
 
\newpage 
 
\setcounter{footnote}{0} 
 
\section{Introduction} 
 
In the counts of galaxies, large fluctuations from field to field  
and from author to author, both in faint  
and bright counts, and in different spectral bands,   
have been reported (e.g. Shanks et al 1989, Tyson 1988,  Cowie et al. 1990,  
Maddox et al. 1990,  Metcalfe et al., 1991,  
Picard 1991, Weir et al. 1995,  
Bertin \& Dennefeld 1997,  
Arnout et al 1997). 
%\cite{tyson88,sh89,tj81,co90,mad90,picard91,bd97,arnout97,metcalfe91}.  
These fluctuations can be as large as a factor of two.  
There have been controversy as to whether these fluctuations  
are due to real clustering or to differences  
in the magnitude zero point of the various surveys.  
Hence, in order to avoid possible systematic errors, it is very important  
to understand the nature of fluctuations  
in a given field of a single survey, once the magnitude system  
and zero point have been carefully calibrated. 
%\cite{picard91,bd97,weir95,arnout97}. 
It is, in fact, possible that discrepancies among these surveys  
are not due mostly to differences in photometric systems  
or in data reduction effects, but rather to real effects,  
i.e. large scale structures.   
In this Letter we propose a method to verify this latter  
possibility in
the actual data. The slope 
and the amplitude of the counts 
are shown to be compatible with a fractal 
distribution of galaxies, and  
we point out  that fundamental information 
about clustering can obtained  by  
  studying  {\it the fluctuations of counts  
as a function of apparent magnitude}.

\section{Average number counts in a  fractal distribution}  
 
As suggested in Baryshev (1981),  
%\cite{yuri}, 
and proposed in a series of papers  
(Sylos Labini et al. 1996, Montuori et al. 1997, Sylos Labini  
Montuori \& Pietronero 1998) 
%\cite{slgmp97,mont97,slmp98}, 
number counts  
versus apparent magnitude can be  
used to test   whether the  
large scale distribution of galaxies  
can be {\it compatible} with a fractal or 
with an homogeneous behavior.  
In this context, we discuss
 the case in which the joint space-luminosity 
distribution $\nu(\vec{r},L)$ can be factorized as the product 
of the number spatial density $n(\vec{r})$ and the luminosity  
function $\phi(L)$  
\footnote{which we take to be Schechter like (Schechter 1976) 
%\cite{sch74},  
but its actual function does not change the  
final results} (Binggeli, Tammann \& Sandage 1988): 
\be 
\label{e1} 
\nu(\vec{r},L) d^3r dL = 
n(\vec{r}) \phi(L) d^3r dL  \,. 
\ee 
This is known to be 
a good approximation in the case of small redshift ($z \ll 1$). 
All the eventual corrections to Eq.~\ref{e1}
(space geometry, K-corrections, 
evolution, etc.) are in fact proportional to $z$ (Yoshii \& Takahara 
1988, Sandage 1995). 
%\cite{sandage,mod1}.

In the case of a 
fractal distribution, the average density seen from a galaxy 
(averaged over enough many observing galaxies) can be written as  
$\left<n(\vec{r})\right>\equiv \Gamma(r)=B r^{D-3}$  
(Pietronero 1987, Sylos Labini et al. 1998) 
%\cite{pi87,slmp98}, 
where $D$ is the fractal dimension; then: 
\be 
\label{e1-bis} 
\left<\nu(\vec{r},L)\right> = 
\Gamma(r) \phi(L)  = 
B r^{D-3} A L^{\delta} e^{-\frac{L}{L_*}}\,. 
\ee  
In this case one ends up with a very simple 
relation for the integrated   
counts as a function of apparent flux ($f=L/(4\pi r^2$)), for unit 
of steradian: 
\be 
\label{C1} 
\left<N(>f)\right> = N_{0} f^{-\frac{D}{2} } \; . 
\ee 
By using the transformation between apparent flux 
and magnitude (Peebles 1993) 
%\cite{pee93} 
\be 
\label{V1a} 
f=\frac{L_*}{4\pi (10 \mbox{pc})^2}10^{0.4(M_*-m)}\,, 
\ee 
where $M_*$ is the cut-off of the luminosity function $L_*$ 
in terms of  
magnitude, 
one obtains 
\be 
\label{C1-bis} 
\left<N(<m)\right> = \tilde N_{0} 10^{-\frac{D}{5}m } \; , 
\ee 
and hereafter we denote $\alpha \equiv D/5$. 
 
Note that, from what concerns the average behavior, the case of 
a homogeneous distribution is included in the fractal case with $D=3$. 
Eq.\ref{e1-bis} has been tested (Sylos Labini, Montuori \& Pietronero 1998; 
Joyce, Montuori \& Sylos Labini 1999) 
%\cite{slmp98,joyce99}  
to be a rather good approximation in local redshift surveys. 
Thus, the exponent of the average counts is  simply related to the  
fractal dimension of galaxies in the three dimensional space 
(see also Sandage, Tammann \& Hardy 1972, Peebles 1993). 
%\cite{sth72,pee93}). 
In Eq.~\ref{e1-bis} $A$ is a normalizing constant such that 
\be  
\label{C1a}  
A = \frac{1}{\int_{L_{min}}^{\infty} L^{\delta} e^{-L/L_*} dL} \;, 
\ee 
where $L_{min}$ is the faintest object observed in current surveys. 
Such a lower cut-off, larger than zero is necessary to avoid 
divergences for $\delta \le -1$.  
Therefore, Eq.~\ref{e1-bis} depends on a combination of 
five different parameters which can be independently measured.  
Three parameters  
are related with 
the luminosity function:  the exponent $\delta$, 
the luminosity cut-off $L_*$  
and the lower cut-off $L_{min}$. These three quantities
have been 
measured with good precision in different redshift surveys 
(Binggeli, Sandage \& Tammann 1988, Efsthatiou, Ellis \& Peterson 1988). 
%\cite{bin88,efs}. 
The fourth parameter is the fractal dimension $D$ and  
the last one, $B$, is
the absolute normalization of the fractal 
distribution. This latter can be, for example, 
defined as the average number of galaxies 
of any luminosity as seen by an average 
observer in a ball of radius $1 \hmp$ 
and can be measured in redshift surveys 
(see Sylos Labini, Montuori \& Pietronero 1998, Joyce, 
Montuori \& Sylos Labini 1999 
%\cite{slmp98,slm98,joyce99}  
for a more detailed 
discussion of the subject).   
 
The amplitude $N_0$ in Eq.\ref{C1} is  given by  
\be 
\label{C2} 
N_0 =   \frac{AB}{2(4\pi)^{ \frac{D-2}{2} } } L_*^{\delta + \frac{D+2}{2}}  
\Gamma_e\left(\delta+ \frac{D}{2}\right)\,, 
\ee 
where $\Gamma_e$ is the Euler's Gamma function. 
 
In view of Eq.\ref{e1-bis} and Eq.\ref{V1a},
 one can compute the average redshift 
of a galaxy with apparent magnitude $m$.
% =-2.5 \log(f) +const$. 
We obtain 
\be 
\label{C2a} 
\langle z \rangle = \frac{h}{3\cdot 10^8}  
\frac{\Gamma_e\left(\frac{D+3}{2} + \delta \right)} 
{\Gamma_e\left(\frac{D+2}{2} + \delta \right)} 
10^{0.2(m-M_*)} \; , 
\ee 
where $h$ is the normalized Hubble's constant. 
 
From  current data both the amplitude and the slope 
of   counts can be estimated. 
In general in the standard  $B_J$ photometric system\footnote{We adopt  
hereafter the standard Johnson-Cousins system 
following the choice  of Arount et al (1997)  
and of  
Bertin \& Dennefeld (1997).}, and in the range of  
magnitude from $\sim 11^m$ to $\sim 19^m$ 
(see Tab.\ref{tab1}), one has $\alpha = 0.50 \pm 0.04$  
corresponding to $D= 2.5 \pm 0.2$. 
The corresponding range of average redshift (Eq.~\ref{C2a}) is  
$ 10^{-3}\ltapprox \langle z \rangle \ltapprox  10^{-1}$. 
Note that, in the faint end part of   counts, 
where the cosmological corrections are known to be relevant, 
the slopes are consistent with the bright end (see Tab.\ref{tab1a}). 
  
%\cite{slgmp97,mont97,slmp98},  
Such a value of $\alpha$ (and hence of $D$) 
is slightly larger than the  value
of $D$  found  
in nearby redshift surveys, which is $D= 2.1 \pm 0.1$ 
up to $\sim 30 \div 50 \hmp$. 
Whether such a difference is due to an increase of fractal 
dimension with scale or it is related to some systematic 
effects in the counts will be discussed in forthcoming papers 
(e.g. Gabrielli \& Sylos Labini 2000). 
%\cite{gsl00,sld00}. 
It is worth to note that Teerikorpi et al. (1988) 
%~\cite{tee98} 
have found a dimension $D=2.35\pm 0.05$ up to 
$\sim 100 \hmp$ by counting galaxies  in real space  
and in volume limited samples.

Note that in the range of $\langle z \rangle \ll 1$ one
expects eventual cosmological and luminosity evolution
corrections to be negligible. However, we propose a further test
to discriminate the importance
of these effects.
 
More specifically, we
 propose   to study in detail the fluctuations
 around the average behavior  of number  
counts as a function of apparent magnitude. 
In fact, as   shown below, this test can discriminate between  
the fractal or smooth cosmological 
nature of the deviation of the $\alpha$ exponent 
from Euclidean behavior ($\alpha = 0.6$).
This study is motivated by the fact that
through number counts we 
can analyze much larger space volumes than in redshift 
surveys. 
In fact, the deepest actual red-shift surveys  
where the fractal dimension has been estimated  
(Joyce et al., 1999)  
%\cite{joyce-2_99} 
  contains  some thousand galaxies, whereas 
magnitude limited surveys can have as many as some millions of galaxies 
up to very faint magnitudes and deep scales (e.g. POSS-II). 
 
\section{Fluctuations}  
 
A very illustrative and simple case is a poissonian  homogeneous
distribution of galaxies. 
In this case the difference between the number  
of points in two equal non overlapping 
volumes is of the order of the square root of the average number.   
The variance of counts can be easily computed from the probabilistic  
definition of Poisson distribution $n(\vec{r})$ and by using again  
Eq.\ref{e1},  obtaining
\be 
\label{V1} 
\sigma_m^2 = \frac{\langle(N(<m)- \langle N(<m) \rangle)^2\rangle} 
{\langle N(<m) \rangle^2} \sim 10^{-0.6 m} \; , 
\ee 
where $N(<m)$ is the number of galaxies with apparent magnitude  
brighter than $m$.  
The average $\langle N(<m) \rangle$  
is given by Eq.\ref{C1-bis} with $D=3$.
A more rigorous derivation,  
considering three point correlation function, can be found 
in Gabrielli \& Sylos Labini (2000). 
%\cite{gsl00}. 
Thus, in the poissonian case,   relative fluctuations   
decrease  exponentially at faint magnitudes. The  
pre-factor in Eq.\ref{V1} 
is simply related to a combination of the parameters 
in Eq.\ref{e1-bis}. 
 
In a fractal distribution the  
typical fluctuation of the number of points $N(r)$ 
in a sphere of radius $r$, with respect to the average value over 
different observers $\langle N(r) \rangle$, 
is always of the same order of the average number  
(e.g. Mandelbrot 1977): 
%\cite{man77}: 
\be 
\label{F0} 
\delta N(r) = \sqrt{ \langle( N(r) - \langle N(r) \rangle)^2 \rangle} \sim  
 \langle N(r) \rangle \; . 
\ee 
This property is very important  
for  counts, which are not averaged over different 
observers (Sylos Labini Montuori \& Pietronero 1998). 
%\cite{slmp98}. 
Eq.~\ref{F0} means that, in a fractal, at any scale, one expects  
to find a void or a structure, the extension of
which is of the same order of the scale itself: 
this is the source of geometrical self-similarity. 
This property implies that  fluctuations in the number of points 
(differential or integral) 
should be, in absolute value,  
always proportional to the average number itself 
and never decreases with distance.  
 
From Eq.~\ref{F0} and Eq.~\ref{e1} one obtains  
that the relative fluctuation in the counts 
as a function of apparent magnitude has  a constant amplitude:   
%\cite{gsl00}:  
\be  
\label{F1}  
\sigma_m \sim const.>0 \,. 
\ee  
Eq.~\ref{F1} describes the ``persistent'' character of fluctuations  
in number counts induced by the fractal 
nature of the spatial distribution.
The numerical value of $\sigma_m$ depends now  
on the same parameters in Eq.\ref{e1}, 
and on some other morphological  
characteristics of the specific studied fractal. 
In fact, 
fluctuations are characterized by higher order correlation functions 
(Blumenfeld \& Ball 1993; 
Gabrielli, Sylos Labini \& Pellegrini 1999) 
%\cite{slmp98,gslp99,ball}  
and the fractal dimension does not 
determine  them univocally. Note that $\sigma_m$ can  be also
very small: its striking feature being in fact 
that it is constant as a function of $m$, and not its absolute amplitude.  
By using simple approximations, it is possible to relate the constant $\sigma_m$
to 
three point correlation function of the distribution (Gabrielli \& Sylos  
Labini 2000). 
In a deterministic fractal, 
  fluctuations have a nearly constant amplitude
  with a log-periodical modulation 
%\cite{sornette} 
(Sornette, 1998) as a function of scale,  because the algorithm 
generating  such a structure is a deterministic one. 
In the more realistic case of stochastic fractals, 
the oscillations are in general a superposition 
of waves, which are periodic in log-space, 
but which have different frequencies and amplitudes. 
 
The poissonian case describes also the situation in which 
one has a spatial distribution of galaxies with a small crossover scale  
$\lambda_0$ to homogeneity and a finite correlation length  
$r_c$ (Gaite et al. 1999, Gabrielli, Sylos Labini \& Durrer 2000). 
%\cite{gaite99,gsld00}. 
A different situation occurs in the case of a spatial distribution 
with a finite homogeneity scale, but an infinite correlation length. 
This case can be thought as obtained by a superposition of a fractal  
distribution to a dominating flat constant density. 
In this case $\left<N(<m)\right>$ is again given by Eq.~\ref{C2}  
with $D=3$ (i.e. it is dominated by the flat constant distribution).  
On the other hand the absolute fluctuation 
$\left<(N(<m)-\left<N(<m)\right>)^2\right>$ is dominated by the fractal 
scale invariant correlations. 
%\cite{gsl00}. 
Consequently, the normalized varaince $\sigma_m^2$ 
is again an exponentially decreasing function of $m$, even  
if with a slower behavior than a poissonian distribution
\be  
\label{F2}  
\sigma_m^2 \sim 10^{(-0.2(3-D)m)} \, , 
\ee  
where $D<3$ is the  
dimension of the fractal superimposed to the constant density.  
 
Note that,  
in general one can have
more complex situations, but the case described  
by Eq.~\ref{F1} is an unambiguous indication of persistent and scale  
invariant 
real space fluctuations typical of statistically self similar irregular  
distributions.

\section{Discussion and Conclusions}

Up to $B_J \sim 18^m$ (i.e.  
$\langle z \rangle <   0.1 $) 
cosmological  models, in the framework of the   
Friedmann solutions, predict an exponent   
$\alpha=0.6$. This is because, one assumes  
an homogeneous distribution starting at very 
small scale, i.e. $5 \div 20 \hmp$  
(Peebles 1993, Davis 1997, Wu, Lahav \& Rees 1999).
%\cite{pee93,dav97}).   
Evolutionary effects, as other  
cosmological corrections, become   
efficient at $z \simeq 1$ (Yoshii \& Takahara 1989, Sandage 1995). 
%\cite{mod1}.  
Such a situation is nearly independent on the value  
of $q_0$, the amount of K-corrections,  
the possible evolution of galaxies   
with redshift, and the photometric band chosen.  
It is important to note  that $N(m,q_0)$ is degenerate to $z$  
in first order: i.e. it is independent on $q_0$ for small 
redshift. 
% \cite{sandage}.  
For instance, 
at  $\langle z \rangle = 0.1$ (i.e. $B_J \simeq 18^m$) 
the deviation from the Euclidean $\alpha = 0.6$ slope 
is less than $10 \%$ for any value of $q_0$. 
Moreover, the slope at fainter magnitudes should be 
{\it a rapidly varying function of the magnitude} 
itself. Clearly, this is not   the case  
for the data shown in Tab.\ref{tab1} and Tab.\ref{tab1a}.

From  an experimental point of view,  
we propose to study the fluctuations 
with respect to the average in the integrated  
number counts $N(<m)$ instead of in the differential 
one $N(m)$, in order to avoid problems with shot noise 
%and poissonian fluctuations  
in magnitude bins.  
In such a way it is clear that  
at bright magnitudes  $\sigma_m$  shows an initial  
decay due to the paucity of bright galaxies in small 
solid angle fields. 
Then, after the integrated number of points has reached 
a large enough value, one should be able to 
detect only the effect of eventual  intrinsic fluctuations. 

However, calibration errors or other 
systematic field-to-field
possible biases can affect the measurement. 
For this reason we suggest  first to 
focus the study to a single sky field at time, instead 
of considering fluctuations in different sky fields. 
That is, after having determined the best fit as in Eq.~\ref{C1}
(or by adding eventual cosmological corrections)
one can study fluctuations in a single 
well calibrated sky field.  
The instrisic nature of 
fluctuations reveals as soon as the shot noise 
contribution becomes negligible. Clearly, there is
a transient between the shot noise regime
and  the instrinsical  fluctuations, where the 
 there is a combination of these two effects.
The range of magnitudes of such a 
combination of shot noise
and intrinsic fractal fluctuations, depends not
only on the number of points, but also
on the solid angle of the survey (see Sylos Labini,
Montuori \& Pietronero 1998 for a more detailed
discussion about the shot noise effect
on the galaxy counts).

It is important to note that the presence of eventual persistent and 
scale-invariant fluctuations, 
in the  $\log N(<m)$ vs. $m$ plot,  
cannot be due to any smooth correction to 
the data as cosmological 
and evolution effects, but they can be the outcome exclusively of 
strongly correlated fractal 
fluctuations.  
The reason  being that smooth linear 
corrections
are  not  able   to produce persistent scale-invariant 
fluctuations on $N(<m)$ 
of the same order of $N(<m$) itself.

It is important to note that
the exponent of the counts can be very sensible
to the photometric band chosen, due to the different
K-correction and kind of objects selected.
However, it is important to stress that the nature
of  fluctuation  structures 
  must be the same in all  different photometric bands. 
In other words, since these fluctuations are intrinsic,  
then they should not depend on the photometric band used. 
On the contrary other possible intervening effects, 
like galactic extinction fluctuations, strongly 
depend on the photometric band chosen.

From the present discussion an important challenge for  
the new generation of experiments is represented  
by  the following questions: 
{\it (i)} Why the slope and the amplitude of the counts  
remain nearly constant beyond $ B_J \approx 18^m$? 
{\it (ii)} Why there is no clear sign 
of change of slope due to galaxy evolution, space-time geometry effects 
and K-correction, even when the average redshift becomes 
to be of order unity ?  
{\it (iii)} The last question concerns the detection of   
persistent and scale-invariant
 fluctuations in the counts.
The new generation of redshift surveys like SSDS 
%\cite{sloan}  
and 2dF, 
%\cite{2df},  
together with the new POSS-II (Djorgovski et al., 2000) 
%\cite{possii} 
photometric survey, will be able to answer 
to these fundamental questions.

\section*{Acknowledgements} 
 
F.S.L. warmly acknowledge  continuous  
and detailed comments 
of Y.V. Baryshev. 
We are grateful to E. Bertin for illuminating 
discussions about data. 
We also thank  H. Di Nella, G. Djorgovski,  
R. Durrer, J.-P. Eckmann,  
M. Joyce, G. Mamon, M. Montuori, G. Paturel, L. Pietronero 
and P. Teerikorpi 
for very useful comments and  suggestions. 
This work is partially supported by the  
 EC TMR Network  "Fractal structures and  self-organization"   
\mbox{ERBFMRXCT980183} and by the Swiss NSF.

\newpage  
 
 \begin{table} 
 \caption{\label{tab1} Determination of the  
differential galaxy 
number counts in the bright end by various authors. In the first column 
it is reported the reference to the original paper, 
in the second the value of the slope and in the third 
the range of $B_J$ magnitudes (Standard Johnson system  
as in Arnouts et al., (1997)) in which the fit has been  
performed. For the transformation between  
different magnitude systems to the standard  
$B_J$ band see Arnouts et al. (1997). For the exponents of the CGCG
(Zwicky et al. 1961-68) and the data of Bertin \& Dennefeld (1997)
see discussion in Gabrielli \& Sylos Labini (2000)} 
 \begin{center} 
 \begin{tabular}{|c|c|c|} 
 \hline 
                      &                  &           	    \\ 
 Author               & $\alpha$         & $\Delta B_J$       \\ 
                      &                  &                  \\ 
\hline		  	   
Bertin \& Dennefeld (1997) & $0.50  \pm 0.01$ & $15.0 \le B_J \le 20.0$\\ 
Maddox et al.  (1990)      & $0.52  \pm 0.01$ & $17.3 \le B_J \le 20.8$\\ 
Weir et al. (1995)         & $0.49  \pm 0.01$ & $17.5 \le B_J \le 20.5$\\ 
CGCG	                   & $0.50  \pm 0.01$ & $11.0 \le B_J \le 15.0$\\ 
%LEDA			   & $0.49  \pm 0.02$ & $11.0 \le B_J \le 14.4$\\ 
Rousseau et al. (1994)     & $0.49  \pm 0.02$ & $11.0 \le B_J \le 15.5$\\ 
\hline 
\end{tabular} 
\end{center} 
\end{table} 
 
\newpage 
 
\begin{table} 
 \caption{\label{tab1a} Determination of the  
slopes of differential galaxy 
counts in the faint end by various authors (see Tab.1).  
The data of Picard (1991) are in the $r$ band 
and the transformation is $B_J \approx r +0.93$.} 
 \begin{center} 
 \begin{tabular}{|c|c|c|} 
 \hline 
                      &                  &           	    \\ 
 Author               & $\alpha$         & $\Delta B_J$       \\ 
                      &                  &                  \\ 
\hline		   
Tyson          (1988)      & $0.45  \pm 0.02$ & $18.0 \le B_J \le 28.0$\\ 
Lilly          (1991)      & $0.38  \pm 0.02$ & $23.0 \le B_J \le 26.5$\\ 
Metcalfe et al.(1991)      & $0.49  \pm 0.05$ & $21.0 \le B_J \le 25.0$\\ 
Metcalfe et al.(1995)      & $0.49  \pm 0.05$ & $24.0 \le B_J \le 27.0$\\ 
Arnouts et al. (1997)      & $0.46  \pm 0.02$ & $20.5 \le B_J \le 24.5$\\ 
Driver et al.  (1994)      & $0.44  \pm 0.02$ & $23.5 \le B_J \le 26.0$\\ 
Picard         (1991)      & $0.45  \pm 0.01$ & $16.0 \le r   \le 19.0$\\ 
\hline 
\end{tabular} 
\end{center} 
\end{table} 
 
\newpage

\end{document}